\documentclass[aps,prl,showpacs,showkeys,twocolumn]{revtex4} 

\usepackage[bookmarks=true,pdftex, unicode=true]{hyperref}
\usepackage{graphicx}
\usepackage{makecell}
\usepackage{amsmath}
\usepackage{esint}
\usepackage{amssymb}
\usepackage[makeroom]{cancel}
\usepackage{color}

\usepackage{hyperref}

\usepackage{xcolor}
  
\def\ket#1{\left|\,#1\,\right\rangle}
\def\bra#1{\left\langle\, #1\,\right|}

\def\mean#1{\langle\,#1\,\rangle}

\def\opone{\leavevmode\hbox{\small1\kern-3.8pt\normalsize1}}

\def\abs#1{\left |\,#1\,\right |}

\newcommand{\fla}[1]{\begin{flalign}#1\end{flalign}}
\newcommand{\bea}{\begin{eqnarray}}
\newcommand{\eea}{\end{eqnarray}}
\newcommand{\bma}{\begin{subequations}}
\newcommand{\ema}{\end{subequations}}
\newcommand{\bwt}{\begin{widetext}}
\newcommand{\ewt}{\end{widetext}}

\newcommand{\avg}[1]{\left\langle #1\right\rangle}

\newcommand{\lra}\simeq
\newcommand{\eeqref}[1]{Eq.~(\ref{#1})}

\newcommand{\de}[0] {{\rm d}}

\begin{document}

\title{Darkness of two-mode squeezed light in  $\Lambda$-type atomic system}

\author{E. S. Moiseev$^{1,3}$}
\email{emoiseev@ucalgary.ca}
\author{Arina Tashchilina$^{1}$}
\author{S. A. Moiseev$^{2}$}
\author{A. I. Lvovsky$^{1,4,5}$}
\email{lvov@ucalgary.ca}

\affiliation{$^1$Institute for Quantum Science and Technology, University of Calgary, Calgary AB T2N 1N4, Canada}
\affiliation{$^2$Kazan Quantum Center, Kazan National Research Technical University, 10 K. Marx, Kazan, 420111, Russia
}
\affiliation{$^3$Department of Theoretical Physics, Institute of Physics, Kazan Federal 	University, 16a Kremlyovskaya St., Kazan, 420111, Russia}
\affiliation{$^4$P. N. Lebedev Physics Institute, Leninskiy prospect 53, Moscow 119991, Russia}
\affiliation{$^5$Russian Quantum Center, 100 Novaya St., Skolkovo, Moscow 143025, Russia}

\begin{abstract}
We show that, under certain circumstances, an optical field in a two-mode squeezed vacuum (TMSV) state can propagate through a lossy atomic medium without degradation or evolution. Moreover, the losses give rise to that state when a different state is initially injected into the medium. Such a situation emerges in a $\Lambda$-type atomic system, in which both optical transitions are driven by strong laser fields that are two-photon resonant with the respective signal modes. Then the interactions of the two signal modes with the ground-state atomic coherence interfere destructively, thereby ensuring the preservation of the TMSV with a particular squeezing parameter. This mechanism permits unified interpretation of recent experimental results and predicts new phenomena.
\end{abstract}
\maketitle
\textit{Introduction.} It has been known since first years of quantum optics that nonclassical properties of optical states, such as squeezing, antibunching, and entanglement, are vulnerable to attenuation \cite{Leonhardt}. Propagating through an attenuator (a lossy channel), the quantum features of an optical state are shared with the environment, and lost when the environment is traced over. Hence it has been a long standing effort to minimize the amount of losses in the preparation and manipulation of these states in order to enhance their utility for quantum information processing \cite{Braunstein05}, quantum metrology \cite{Giovanetti06}, and other applications. 

In this paper, we challenge this paradigm, showing a family of nonclassical, entangled states of light that not only propagate through an attenuating medium without being affected by losses, but, moreover, are created thanks to these losses. That is, any other state, after entering and propagating through this medium, is converted into such a state. We call these states optical dark (OD) states, in analogy to the dark states of atoms which do not absorb light in spite of it being in resonance with the atomic transition. 

\begin{figure}
	\includegraphics[width=0.22\textwidth]{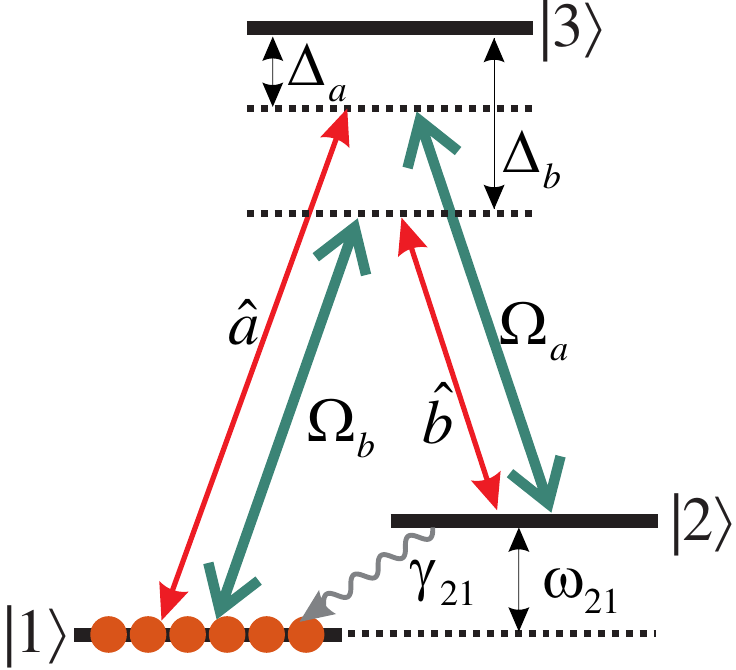}
	\caption{The atomic level scheme. The two  quantum modes $\hat a$ and $\hat b$ are in two-photon resonance with the strong classical fields $\Omega_a$ and $\Omega_b$, respectively. When the quantum modes are in the two-mode squeezed vacuum state with a certain degree of squeezing, they do not interact with the atomic medium during propagation.
	}
	\label{scheme}
\end{figure}

Similarly to the atomic dark state, the OD state arises in $\Lambda$-shaped atomic systems. The two ground states are coupled to each other in a Raman-like manner by two pairs of fields. In each pair, one field is quantum and the other is a strong laser (Fig.~\ref{scheme}). In this way, the quantum fields directly interact with the atomic ground states: absorption of a photon in mode $\hat a$ transfers a photon from level $\ket 1$ to level $\ket 2$, whereas mode $\hat b$ has the opposite effect. When both modes are populated with photons, these processes occur in superposition. Moreover, if the state of these modes is TMSV with a certain squeezing parameter (determined by the ratio of the effective  coupling constants between the optical modes and matter), the two processes interfere destructively, thereby effectively precluding the interaction of the atomic and optical states. Then, even if the ground state coherence experiences decay, this OD state will propagate through a gas of such atoms without any loss or evolution. 

The physics of the phenomena studied here are closely related to those of \cite{Mushik2011,PhysRevLett.107.080503}, where entanglement of two macroscopic atomic ensembles has been driven created by dissipative phenomena. In fact, it is the same processes that generate the entangled states of both light and atoms, as we show below. 

We emphasize the difference between the OD-state setting studied here and the four-wave mixing regime which is known to produce two-mode squeezing in a system similar to that of Fig.~\ref{scheme}. In four-wave mixing, the Raman population transfer between the ground states is eliminated by either working away from the two-photon resonance \cite{Guzman2006,Marino09,MacRae2012}, or by means of  electromagnetically induced transparency \cite{Kolchin2007}. In this case, the atomic state is decoupled from the evolution, resulting in the usual two-mode squeezing Hamiltonian, which leads to exponential growth of squeezing as the field is propagating through the sample (albeit with fragility to losses). We, in contrast, work under the conditions of two-photon resonance, so both quantum fields, taken individually, experience significant Raman absorption or amplification. In this regime, the squeezing is not amplified with the propagation. It stays constant, but any other state of the two-mode field asymptotically approaches the OD state thanks to that Raman interaction.

\paragraph{Concept.} We limit our analysis to one dimension and assume that the light fields propagate along $z$ direction. The atomic ensemble has the length $L$ and linear atomic density $n_0={N}/{L} $, where $N$ is the total number of atoms. The atoms, initially prepared in the state $\ket 1$, are described by slowly-varying collective coherence operators $\hat{S}_{nm}(z)= \sum_{j=1}^{N} \hat{S}_{nm}^j\delta(z-z_j)$ with the commutation relation $\left[ \hat{S}_{nm}(z'), \hat{S}_{mn}(z'') \right]=n_0\delta(z'-z'') $, where $\hat{S}_{nm}^j=\ket{n_j} \bra{m_j}$ for the $j$'th atom.

The two quantum fields, which we call signal and idler, are described by annihilation operators $\hat{a}$ and $\hat{b}$, whose commutation relation is $[\hat{a}(z),\hat{a}^{\dagger}(z')]=[\hat{b}(z),\hat{b}^{\dagger}(z')]=\delta(z-z')$. 
The Rabi frequencies of the corresponding control fields are $\Omega_a$ and $\Omega_b$, respectively.

The interaction Hamiltonian in the rotating-wave approximation is then  \cite{GorshkovPRA2011,Fleischhauer2013}:
\fla{
\hat{H} =  \hbar\int_{0}^{L} dz \bigg\{  \left(g_{31}  \hat{a}(z) e^{-i\omega_a t} + \Omega_b e^{-i\omega_{\Omega_b} t}\right) \hat{S}_{31}(z) 
\nonumber \\
+ \left( g_{32} \hat{b}(z) e^{-i\omega_b t} + \Omega_a e^{-i\omega_{\Omega_a} t}\right)\hat{S}_{32}(z) +{\rm H.c.} \bigg\}
}
where $g_{31}$ and $g_{32}$ are  photon-atom coupling constants  for the corresponding optical transitions \cite{HammererReview}, $\omega_{a,b}$ and $\omega_{\Omega_{a,b}}$ are the carrier frequencies of the quantum and control fields, respectively. We assume the phase matching condition to hold, so the relative phase of the atomic and optical operators stays constant throughout the sample. 

If the signal and control fields are far detuned for the respective atomic transitions (i.e., $\Delta_{a,b}\gg \gamma_{3},\Omega_{a,b}$, where $\gamma_3$ is the spontaneous decay rate from the excited level $\ket3$), we can adiabatically eliminate level $\ket 3$, arriving at the following effective interaction Hamiltonian:
\fla{
&\hat{V}_{\text{eff}} = \hbar \int^{L}_{0} dz\left( g_a^{*}\hat{a}^{\dagger} + g_b\hat{b}\right)\hat{S}_{12}+{\rm H.c.} ,
\label{eff_Hamiltonian}
}
where $g_a=\frac{g_{31}\Omega_a^{*}}{\Delta_a}$ and $g_b=\frac{g_{32}\Omega^{*}_i}{\Delta_b}$ are the effective coupling constants of the signal and idler modes with the spin wave (we specialize to the case $|g_b|<|g_a|$). Equation \eqref{eff_Hamiltonian} is valid if the respective control and quantum field pairs are in a two-photon resonance with the ground states that are ac Stark shifted by the control fields, which we assume to be the case. Another important assumption is that the overwhelming majority of the atomic population is in state $\ket 1$, which is valid on time scales that are small compared to the inverse optical pumping rate associated with the control field $\Omega_b$: $\frac{\abs{\Omega_b}^2 \gamma_{3} }{\Delta^2_b}t\ll1$. In this case, the Hilbert space associated with the atomic ground state coherence becomes isomorphic to that of the harmonic oscillator under the Holstein-Primakoff transformation \cite{HolsteinPrimakoff1940}

To demonstrate the OD state, we perform a Bogoliubov transformation of the signal and idler modes according to 
\begin{equation}\label{bogo}
\hat{B}= \alpha^{-1}_0 (\hat{a}+\epsilon\hat{b}^{\dagger}),\ \hat{D}= \alpha^{-1}_0 (\hat{b}+\epsilon\hat{a}^{\dagger}), 
\end{equation}
where $\alpha_0=\sqrt{1-\abs{\epsilon}^2}$ and $\epsilon=g_b/g_a<1$ (hereafter we assume the phase convention for $\hat a$ and $\hat b$ to be chosen such that $g_a$ and $g_b$ are real and positive). Then the  Hamiltonian (\ref{eff_Hamiltonian}) is transformed into 
\fla{
\hat{V}_{\text{eff}}(t)= \hbar\alpha_0g_a \int^{L}_{0} dz \left(\hat{B}^{\dagger} \hat S_{12} + \hat{B} \hat S_{12}^\dag\right).\label{BrightHamiltonian}
}
We see that the field in the ``dark" mode $\hat{D}$, no matter what state it is in, is decoupled from the interaction ($[\hat{D},\hat{V}_{\text{eff}}]=0$). The atomic system is coupled only to the ``bright" mode $\hat{B}$. If the atomic coherence experiences relaxation, this beam-splitter-like coupling will result in absorption, so the bright mode will gradually decay into its ground state $\ket{0_B}$. If the bright mode is initially prepared in that state, it will propagate through the atomic sample without evolution, akin to a conventional optical mode in the vacuum state propagating through an ensemble of resonant atoms. Therefore any state of the form $
\ket{0_B, \Phi_D},$
with arbitrary $\ket\Phi$, is an OD state. This state, combined with the collective atomic state  $\ket{0_A} = \ket{1_1\ldots1_N}$, is an  eigenvector of the interaction Hamiltonian  (\ref{eff_Hamiltonian}) with eigenvalue 0. 

\begin{figure*}[t]
	\includegraphics[width=0.85\textwidth]{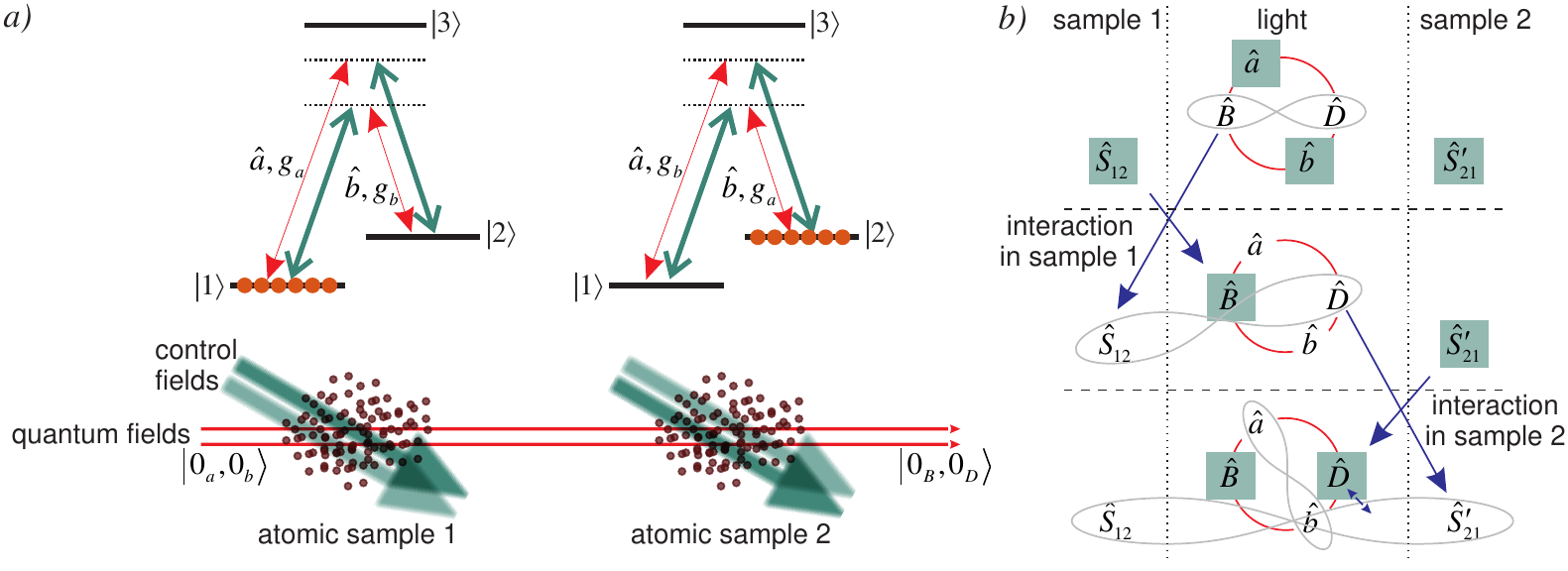}
	\caption{Generating the OD state (\ref{darkstate}) from vacuum input $\ket{0_a,0_b}$. a) Scheme of the experiment and energy level diagrams. The coupling constants and the populations of the two samples are exchanged with respect to the two ground states. b) Exchange of entanglement among the optical modes and atomic coherences. Red circles symbolize the Bogoliubov transformation that relate the pairs of optical modes $(\hat a,\hat b)$ and $(\hat B,\hat D)$.  Shading marks the modes that are in the vacuum state, infinity symbols denote the TMSV state. Top: at the entrance of the first sample, the atomic ensembles and the physical modes $\hat a$ and $\hat b$ are in their ground states, which means that the Bogoliubov modes $\hat B$ and $\hat D$ are TMSV entangled as per \eeqref{vacuum_state}. Center: the interaction in the first sample swaps the optical Bogoliubov mode $\hat B$ and the atomic coherence $\hat S_{12}$. Bottom: in the second sample, the contents of $\hat D$ and  $\hat S'_{12}$ are swapped. Now the atomic samples are TMSV entangled while the Bogoliubov modes are in the vacuum state, which means that the physical light modes are in the TMSV state as well according to \eeqref{darkstate}.
	}
	\label{generation_scheme}
\end{figure*}

\paragraph{Ground OD state.} Of particular interest among the OD states is the vacuum state $\ket{0_B,0_D}$ of modes $\hat{B}$ and $\hat{D}$. Because the original modes ($\hat{a}$, $\hat{b}$) are related to ($\hat{B}$, $\hat{D}$) via the Bogoliubov transformation, the state $\ket{0_B,0_D}$ in the eigenbasis of ($\hat{a}$, $\hat{b}$) is a TMSV:
\begin{align}
\ket{0_B,0_D}&=\exp\left[r(\hat a\hat b -\hat{a}^{\dagger}\hat{b}^{\dagger})\right]\ket{0_a,0_b}\nonumber\\
&=\alpha_0\sum_n \epsilon^n  \ket{n_a,n_b}, 
\label{darkstate}
\end{align}
where $r=\frac12\log\frac{1-\epsilon}{1+\epsilon}$ is the squeezing parameter and $\ket{n_{a,b}}$ denotes number states. This state is characterized by the mean photon numbers $\mean{\hat{n}_a}=\mean{\hat{n}_b}=\epsilon^2/{\alpha_0^2}$ and the position/momentum quadrature correlation 
\fla{
\avg{(X_a\pm X_b)^2}=\avg{(P_a\mp P_b)^2}=e^{\pm 2r}=\frac{1\pm\epsilon}{1\mp\epsilon},
\label{variancesX}
}
with $\epsilon=r=0$ corresponding to the standard quantum limit \cite{SqLightReview}. The squeezing becomes infinite in theory for $\epsilon\to1$.

State \eqref{darkstate} coincides with the vacuum state $\ket{0_a,0_b}$ if $\epsilon=0$. This case corresponds to the idler control field being absent, so the signal field experiences Raman absorption, decaying into the vacuum state while propagating through the sample. On the other hand, for $\epsilon\ne 0$, the physical vacuum is not an OD state. To see this, we notice that this state is two-mode squeezed in the basis of the Bogoliubov bright and dark modes:
\begin{equation}
\ket{0_a,0_b} = \exp\left[r(-\hat{B}\hat{D}+r\hat{B}^{\dagger}\hat{D}^{\dagger})\right]\ket{0_B,0_D}.
\label{vacuum_state}
\end{equation} 
If this state is injected into our atomic sample, the bright mode will be absorbed by the atoms, decaying into $\ket{0_B}$. With the atomic coherence dissipating into the environment, mode $\hat D$ becomes thermal,
with the quadrature variance $\avg{X_D^2}=\avg{P_D^2}=\frac12\cosh 2r=\frac{1+\epsilon^2}{1-\epsilon^2}$: 
\fla{
	\hat{\rho}_{B,D}= \abs{\alpha_0}^2  \ket{0_B}\bra{0_B}\otimes\sum_n \epsilon^{2n}  \ket{n_{D}} \bra{n_{D}}.
	\label{thermal_D}
}
State \eqref{thermal_D}, albeit unpure, is two-mode squeezed in the basis of modes $\hat a$ and $\hat b$ by no more than  a factor of $2$ with respect to the standard quantum limit:
\begin{align}\label{SqBtwSamp}
\avg{(X_a\pm X_b)^2}&=\avg{(P_a\mp P_b)^2}\\&
=\frac12 e^{\mp 2r}(1+\cosh 2r)=(1\pm\epsilon)^{-2}.\nonumber
\end{align}
This squeezing can be experimentally observed by performing a homodyne measurement on the signal and idler modes upon exiting the sample.

\begin{figure}[htb]
	\centering
	\begin{tabular}{@{}c@{}}
		\includegraphics[width=0.4\textwidth]{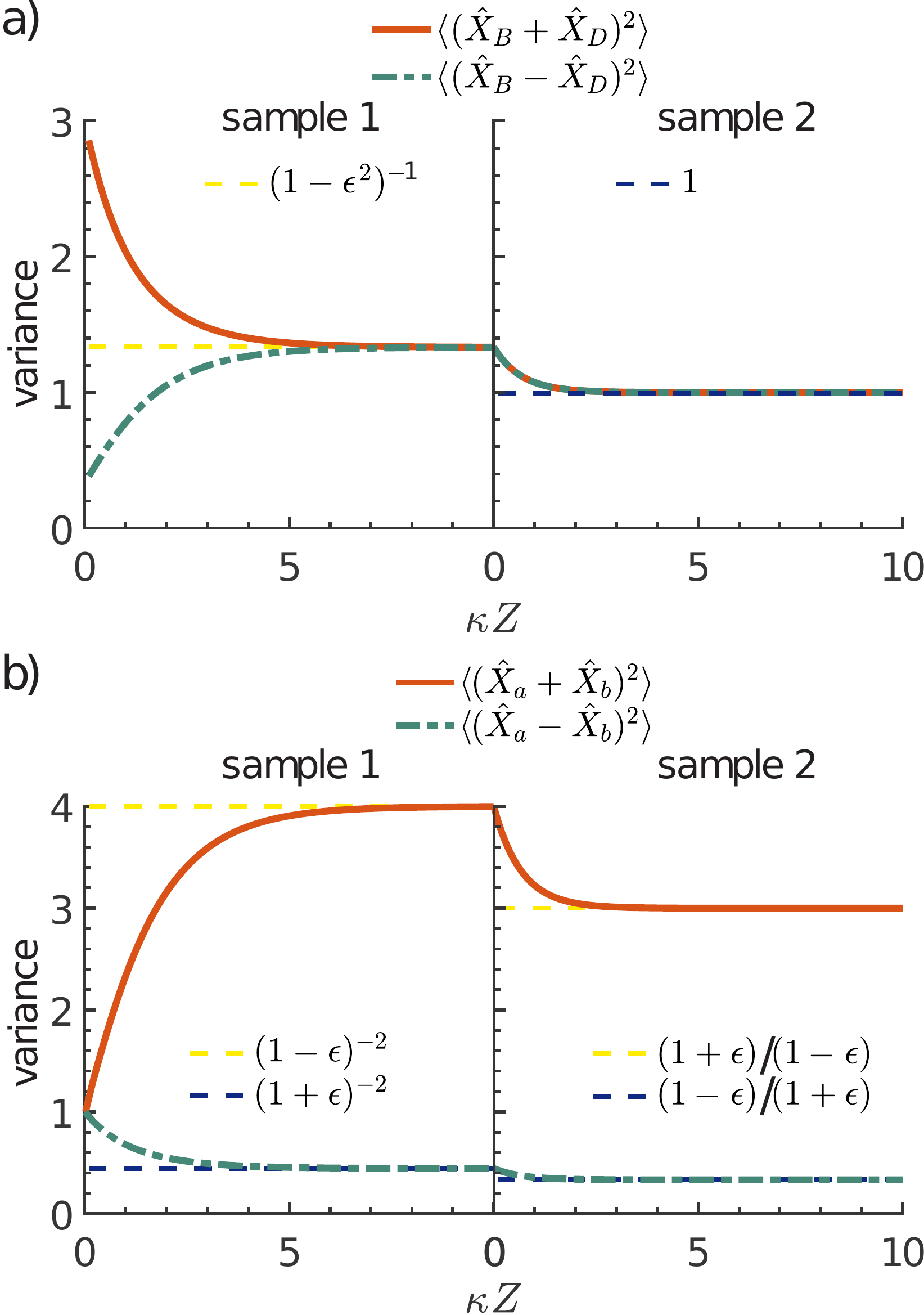}
	\end{tabular}
	\caption{(a) Development of entanglement of the optical  modes as they propagate through the two samples for the vacuum input $\ket{0_a,0_b}$ and $\epsilon=0.5$. a) Bogoliubov $(\hat B,\hat D)$ modes; b) Physical  modes $(\hat a,\hat b)$. The variances of the individual position quadratures as well as their sum and difference are displayed. At the entrance of the first sample, the  Bogoliubov modes are in the TMSV state. Between the samples, the state is mixed and described by Eqs.~\eqref{thermal_D} and \eqref{SqBtwSamp}. After the second sample, the Bogoliubov modes are in the vacuum state, and the physical modes are in TMSV described by Eqs. (\ref{darkstate}) and (\ref{variancesX}) }
	\label{var}
\end{figure}

The fact that an entangled state remains unchanged while propagating through an absorbing medium, while the vacuum state loses its purity and becomes entangled, is highly counterintuitive. In the light of the above discussion, we explain this by observing that the interaction of the light with the environment occurs by way of the bright mode $\hat B$. The pair of modes $(\hat B,\hat D)$ therefore defines the decoherence-preferred basis. States that are entangled in this basis do decohere. However, because this basis is itself entangled in terms of the physical modes $(\hat a,\hat b)$, this decoherence presents itself as growth of entanglement of the latter modes. 

We now show how our scheme can be extended to  prepare the two mode squeezed vacuum state
(\ref{darkstate}) from physical vacuum input. To that end, we  send the optical modes through an additional, similar atomic sample [Fig.~\ref{generation_scheme}(a)] with the atomic population prepared in state $\ket 2$. 
In addition, we invert the ratio $\epsilon$, which is equivalent to exchanging the values of the coupling constants $g_a$ and $g_b$ on the atomic transitions. 
This is done by adjusting the amplitudes and phases of the Rabi frequencies $\Omega_a$ and $\Omega_b$. 
In this case the effective Hamiltonian is
\fla{
	\hat{V'}_{\text{eff}}= \hbar\alpha_0g_a \int^{L}_{0} dz \left(\hat{D}^{\dagger} {\hat S}'_{21} + \hat{D} (\hat S'_{21})^\dag\right),\label{DarkHamiltonian}
}
where the primes mark the second sample. Now mode $\hat D$ becomes bright and experiences absorption, while mode $\hat B$ is dark and does not evolve. Since, after the first sample, mode $\hat B$ is already in the vacuum state, propagation through the second sample will yield the double-vacuum state (\ref{darkstate}) of modes $\hat B$ and $\hat D$.

 We see, remarkably, that a pure entangled two mode  state (\ref{darkstate}) can be not only preserved, but also generated through a dissipative (absorptive) process. This is  a potentially powerful method to generate TMSV, which is the universal starting point for the preparation of arbitrary complex states of light \cite{RaymerReview}. In addition to robustness to losses, our technique permits easy control of the squeezing parameter $r(\epsilon)$ by adjusting the strengths of the control fields.
Potential detrimental factors such as  spontaneous emission from level $\ket{3}$ and nonlinearities caused by a finite population in level $\ket 2$ can be suppressed by reducing the interaction time and working at sufficiently large one-photon detunings.

\paragraph{OD states and quantum optical memory.} The beam-splitter form of coupling defined by the Hamiltonian \eqref{BrightHamiltonian} in an optically deep medium leads to the swapping of the states between the optical mode $\hat B$ and the atomic coherence $\hat S_{12}$, which is the basis of many quantum optical memory protocols \cite{HammererReview,QMReview}. Similarly, the interaction \eqref{DarkHamiltonian} that takes place in the second sample will swap the contents between $\hat D$ and  $\hat S'_{12}$. So far, we assumed the atomic spin state to dissipate after this swap due to the ground state decoherence. However, an interesting interpretation arises if we include the atomic state into our analysis, which is justified if its dissipation is sufficiently slow.

If the state entering the first sample is physical vacuum $\ket{0_a,0_b}$, modes $\hat B$ and $\hat D$ are in the TMSV state \eqref{vacuum_state}. When the light propagates through the atoms, this entanglement will be swapped to the first and second samples [Fig.~\ref{generation_scheme}(b)].  At the same time, modes $\hat B$ and $\hat D$ will now take over the vacuum from the initial atomic states, thereby bringing the  signal and idler modes $\hat a$ and $\hat b$ into the TMSV state  according to  Eq.~\ref{darkstate}. In this way, both the atomic and optical states will become entangled, at the same time remaining in states that are separable from each other. 

These phenomena, in our understanding, offer alternative physical intuition behind the recent experiments on entanglement generated by dissipation \cite{PhysRevLett.107.080503} and producing the TMSV state of light \cite{Wasilewski:09}. In these experiments, the two samples are atomic cesium vapor cells, with the  roles of levels $\ket {1,2}$ played by the magnetic states $\ket{m=-4,-3}$ of the ground level  $6S_{1/2}, F=4$ in one of the samples, and $\ket{m=3,4}$ in the other one. Such a configuration automatically ensures the inversion of the coupling constants $g_a$ and $g_b$ between the two samples.

\paragraph{Propagation of OD states.} To explicitly show that OD state is preserved in an ensemble with incoherent decay, we study the evolution of the modes by taking into account the effective Hamiltonian (\ref{DarkHamiltonian}) with the free Hamiltonians of atoms, dark and bright fields, we find the following Heisenberg-Langevin equations:
\fla{
\left(\partial_t +c\partial_z \right) \hat B &=-i\alpha_0 g_a\hat{S}_{12}, \quad  \left(\partial_t +c\partial_z \right)\hat D =0,  \label{brightEvo} \\
\frac{d \hat S_{12}}{dt} &=-\gamma_{12} \hat{S}_{12}-i\alpha_0 n_0 g_a\hat B
-i\hat{F}_{12}(z,t), \label{coherenceEvo}
}
where $\gamma_{12}$ is the ground state coherence decay constant, $\mean{\hat{F}_{12}}$ are delta-correlated Langevin forces with the well-known correlation functions $\mean{F_{12}^{\dagger}(z,t)}=\mean{F_{12}(z',t')} =0$, $\mean{F_{12}(z,t)F_{12}^{\dagger}(z', t')}=2\gamma_{12}\delta(z-z')\delta(t-t')$.
A general  solution to the above equations can be found similiar to Ref.~\cite{Polzik2000}. 
Using the Fourier transformation  $\hat{B}(\omega,Z)= \frac{1}{\sqrt{2\pi}}\int \hat{B}(\tau,Z)e^{i\omega \tau} d\tau$ in the co-moving reference frame $\tau=t-z/c$, $Z=z$ and parameterizing the optical depth via $\kappa = \alpha_{0}^2 n_0 \abs{g_s}^2/(c\gamma_{12})$, we arrive at:

\begin{widetext}
\fla{
\hat{B}(\omega,Z)  &=  e^{-  \frac{\kappa Z}{1-i\omega/\gamma_{12}}} \hat{B}(\omega,0) + \int_0^Z dZ'\frac{ \sqrt{\kappa/(c\gamma_{12})}}{1-i\omega/\gamma_{12}} e^{\frac{\kappa(Z'-Z)}{1-i\omega/\gamma_{12}}}  \left( \hat{F}_{12}(\omega,Z') + e^{i\omega_0 t}\hat{S}_{12}(t_0,Z') \right).
\label{solFourier}
}
\end{widetext}
Mode $\hat{B}$ exhibits usual Beer's absorption and tends to the vacuum state $\ket{0_B}$ in the limit of infinite optical depth (Fig.~\ref{var}). Solution \eqref{solFourier} allows us to write a closed form expression for the behavior of the position and momentum quadratures of mode $\hat B$  as it propagates through the sample. For example, if  the initial state is $\ket{0_a,0_b}$, we have (Fig.~\ref{var}):
\begin{align*}
\mean{\Delta \hat X_B^2 (\omega, Z)} &= \mean{\Delta \hat P_B^2 (\omega, Z)}\\
&=\left( \frac{1}{2}+\frac{\epsilon^2}{1-\epsilon^2} e^{-\frac{2\kappa Z}{1+(\omega/\gamma_{12})^2}}\right)
\end{align*}
We see that the quadrature variance of the bright field evolves to the value of  $\frac 12$, which is characteristic of the vacuum state.
  
\begin{figure}[b]
\includegraphics[width=0.45\textwidth]{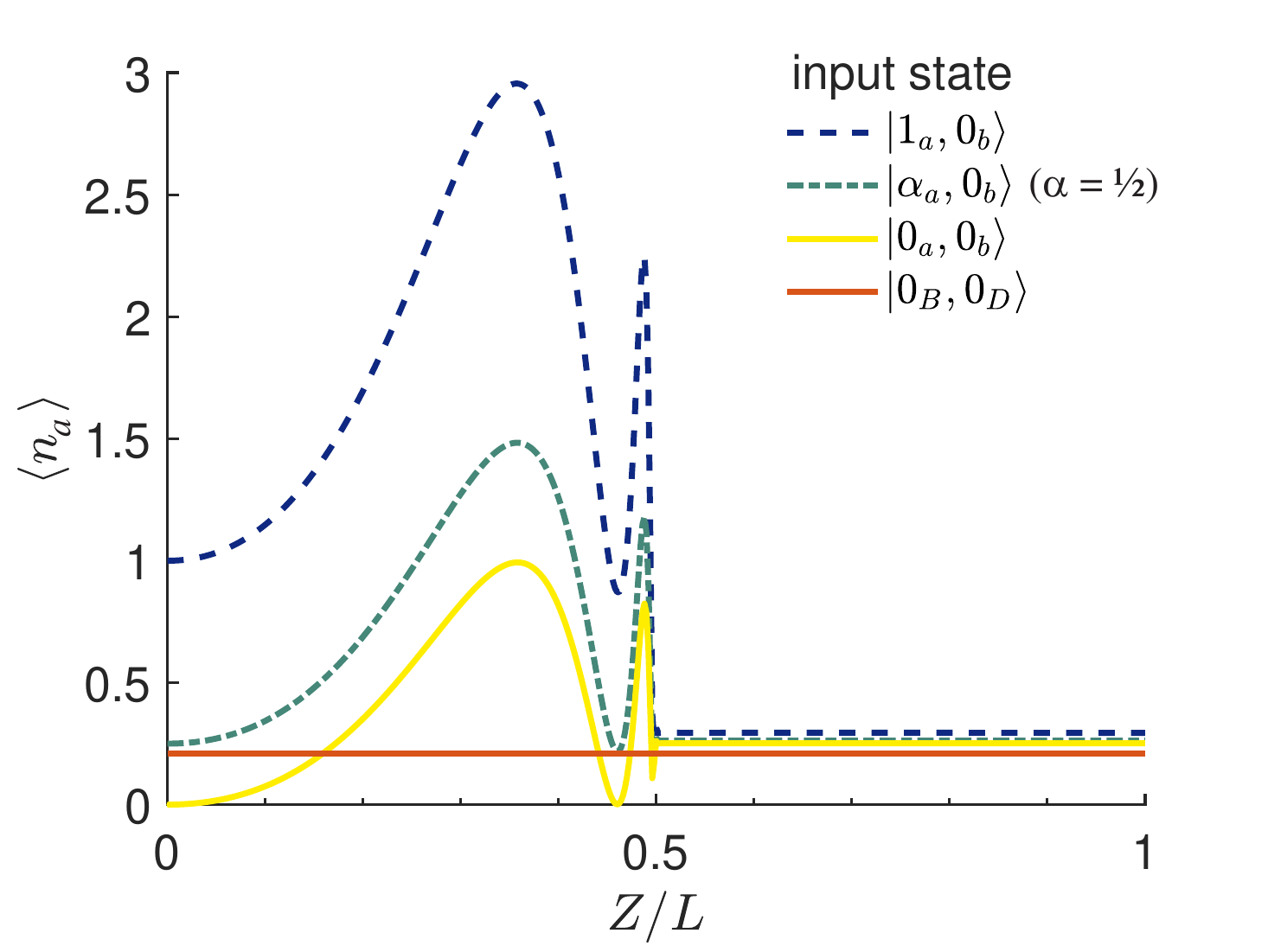}
  \caption{
  Photon number of the signal mode propagating through an atomic sample with a longitudinal inhomogeneous broadening of the ground state transition according to $\omega_{21}(z)=\omega_{21}(0)+\beta (Z-L/2)$, where $\beta$ is gradient  constant.
  The optical depth normalized by the inhomogeneous broadening is $ \kappa \gamma_{12}/\beta =5 $.      
}
\label{GEM}
\end{figure}

It is interesting to analyze the emergence of OD states in the context of gradient echo memory setting \cite{Hosseini2011,higginbottom2015dual}, in which the frequency of the ground state transition varies along the sample. Figure \ref{GEM} shows the the number of photons in the signal mode as it propagates through the sample. When the fields enter the atomic sample, the two-photon detuning for each pair of control and quantum fields is significant, so a four-wave mixing process develops, leading to amplification. At the center of the sample, with the onset of two-photon resonance, the bright mode is absorbed; its optical state  becomes vacuum $\ket{0_B}$. 

Curiously, with further propagation, this state remains unchanged in spite of the reemergence of the two-photon detuning.
This can be intuitively explained as follows. In the presence of two-photon detuning $\delta_{12}(Z)$,  the Hamiltonian (\ref{BrightHamiltonian}) acquires an additional position-dependent term $\int  \delta_{12}(Z) \hat{S}_{22}(Z)\de Z$ \cite{MoiseevNJP2013}. When this detuning is significant, it dominates the light-atom interaction and results in the evolution of the dark field according to the phase shift $B(\omega,Z)= e^{-i\phi(Z,\omega)} B(\omega,0)$ with $\phi(Z,\omega) \propto \frac{\kappa \de Z}{\delta_{12}(Z)-\omega}$. In the Schr\"odinger picture, this phase shift corresponds to the evolution operator  $\hat{U}=e^{-i   \int d\omega \phi(Z,\omega) \hat{B}^{\dagger}(\omega,0)\hat{B}(\omega,0) } $. If the bright mode is in the vacuum state, this operator equals identity, so no evolution is present.

\textit{Conclusion.} We have proposed a new formalism to treat the propagation of quantum light modes that are coupled to the ground state coherence via a two-photon Raman transition mediated by a control field. 
By using the approach, we have found a number of new properties in the off-resonant interaction of entangled light fields with three-level atomic media. 
In particular light entaglement can play significant role for formation of transparency in a similiar fashion as superposition in atomic dark states. 
This formalism stipulates a previously undescribed mechanism of generating optical squeezing that is robust to losses.

The OD state interpretation features a number of paradoxical features and, we hope, will provide a powerful visualization tool for complicated theoretical analysis. The phenomena studied here can be found in other fields of quantum physics, for example, in  optomechanics \cite{PhysRevLett253601}.

\begin{acknowledgments}
We thank E. Polzik for illuminating discussions.
This study was funded by NSERC and CIFAR. SAM's research was supported by the Russian Science Foundation, project no. 14-12-01333-P.
\end{acknowledgments}

\bibliography{QMnoise}

\end{document}